\documentclass[aps, prb, reprint, floatfix, superscriptaddress, longbibliography]{revtex4-2}

\usepackage{amsmath}
\usepackage{graphicx}
\usepackage{color}
\usepackage{tabularx}
\usepackage{xr}

\usepackage{moreverb} 

\usepackage{verbatim}

\makeatletter

\begin{document}

\title{Multi-mode Brownian Dynamics of a Nanomechanical Resonator in a Viscous Fluid}

\author{H. Gress}
\affiliation{Department of Mechanical Engineering, Division of Materials Science and Engineering, and the Photonics Center, Boston University, Boston, Massachusetts 02215, USA \looseness=-1}

 \author{J. Barbish}
 \affiliation{Department of Mechanical Engineering, Virginia Tech, Blacksburg, Virginia 24061, USA \looseness=-1}
 
\author{C. Yanik}
\affiliation{SUNUM, Nanotechnology Research and Application Center, Sabanci University, Istanbul, 34956, Turkey \looseness=-1}
\affiliation{Faculty of Engineering and Natural Sciences, Sabanci University, Istanbul, 34956, Turkey \looseness=-1}

\author{I. I. Kaya}
\affiliation{SUNUM, Nanotechnology Research and Application Center, Sabanci University, Istanbul, 34956, Turkey \looseness=-1}
\affiliation{Faculty of Engineering and Natural Sciences, Sabanci University, Istanbul, 34956, Turkey \looseness=-1}

\author{R. T. Erdogan}
\affiliation{Department of Mechanical Engineering, Bilkent University, Ankara, 06800, Turkey \looseness=-1}
\affiliation{National Nanotechnology Research Center (UNAM), Bilkent University, Ankara, 06800, Turkey \looseness=-1}

\author{M. S. Hanay}
\affiliation{Department of Mechanical Engineering, Bilkent University, Ankara, 06800, Turkey \looseness=-1}
\affiliation{National Nanotechnology Research Center (UNAM), Bilkent University, Ankara, 06800, Turkey \looseness=-1}

\author{M. Gonz\'alez}
\affiliation{Aramco Americas, Aramco Research Center--Houston, Houston, Texas 77084, USA \looseness=-1}

\author{O. Svitelskiy}
\affiliation{Department of Physics, Gordon College, Wenham, Massachusetts 01984, USA \looseness=-1}

\author{M. R. Paul}
\affiliation{Department of Mechanical Engineering, Virginia Tech, Blacksburg, Virginia 24061, USA \looseness=-1}

\author{K. L. Ekinci}
\email{ekinci@bu.edu}
\affiliation{Department of Mechanical Engineering, Division of Materials Science and Engineering, and the Photonics Center, Boston University, Boston, Massachusetts 02215, USA \looseness=-1}

\date{\today}

\begin{abstract}
Brownian motion imposes a hard limit on the overall precision of a nanomechanical measurement. Here, we present a combined experimental and theoretical study of the Brownian dynamics of a quintessential nanomechanical system, a doubly-clamped nanomechanical beam resonator, in a viscous fluid. Our theoretical approach is based on the  fluctuation-dissipation theorem of statistical mechanics: We determine the  dissipation from fluid dynamics; we  incorporate this dissipation into the proper elastic equation to obtain the equation of motion; the fluctuation-dissipation theorem then directly provides an analytical expression for  the position-dependent power spectral density (PSD) of the displacement fluctuations of the beam.  We compare our theory  to  experiments on nanomechanical beams immersed in air and water, and  obtain excellent agreement.  Within our experimental parameter range, the Brownian force noise driving the nanomechanical beam has a colored PSD due to the ``memory" of the fluid; the force noise remains  mode-independent and  uncorrelated in space. These conclusions are not only important for nanomechanical sensing but also provide  insight into the fluctuations of  elastic systems at any length scale. 
\end{abstract}


\maketitle

\section{Introduction}

Brownian fluctuations of mechanical systems have been a topic of active research in physics since the 1920s~\cite{uhlenbeck1930theory, kappler1931versuche}. Early electrometers~\cite{parson1915highly} and galvanometers~\cite{moll1925lxv} that featured proof masses attached to linear springs displayed irregular movements around their equilibrium points despite all ``precautions and shields"~\cite{barnes1934brownian}. These early experiments eventually led to the realization that the observed fluctuations, namely, Brownian motion, were of fundamental nature and limited the overall precision of mechanical measurements~\cite{barnes1934brownian}. A century later, Brownian motion still remains centrally relevant to precision metrology and sensing based on  mechanical systems---in particular, nanoelectromechanical systems (NEMS)~\cite{bachtold2022mesoscopic} and AFM microcantilevers~\cite{butt1995calculation}. These state-of-the-art miniaturized mechanical systems are even more susceptible to Brownian noise than their macroscopic counterparts since they tend to be extremely compliant to forces.  

The Brownian dynamics of a nanomechanical resonator can be  formulated using elasticity theory and statistical mechanics. Elasticity theory provides a dissipationless equation of motion, such as the beam equation. Solving this equation under a harmonic ansatz and subject to boundary conditions maps the dynamics of the nanomechanical resonator onto that of a collection of eigenmodes, i.e., spring-mass systems, with discrete eigen-frequencies and mode shapes  (eigenfunctions)~\cite{saulson1990thermal, cleland2013foundations}.  In the simplest approximation of Brownian dynamics, each eigenmode is assumed to have a constant and spatially uniform dissipation, resulting in a  Brownian force noise that is delta-function correlated in both time and space. These assumptions result in a theoretical expression for the power spectral density (PSD) of the displacement fluctuations of the beam as a sum of the PSDs of  the individual uncorrelated eigenmode  fluctuations~\cite{saulson1990thermal, cleland2002noise}. In the limit of small dissipation, multi-mode noise measurements on   cantilevers~\cite{paolino2009direct}, microdiscs~\cite{wang2014spatial}, microtoroids~\cite{mcrae2010cavity},  nanowires~\cite{gloppe2014bidimensional} and macroscopic elastic systems~\cite{arcizet2006high} all agree with this first-order approximation. 

Most mechanical systems, however, \textit{do} come with some ``memory," making the above-mentioned  assumption of temporally uncorrelated force noise inaccurate \cite{saulson1990thermal}. If the dissipation is spatially non-uniform, i.e., position dependent, the force noise  between different eigenmodes becomes correlated \cite{levin1998internal}, with the eigenmode expansion of the force noise becoming non-trivial. For elastic systems with spatially non-uniform dissipation, alternative approaches to calculate the noise PSD have been developed \cite{levin1998internal, liu2000thermoelastic} and experimentally tested \cite{yamamoto2001experimental, yamamoto2002study, schwarz2016deviation}. 

For a nanomechanical resonator immersed in a viscous fluid, memory  comes from the flow-resonator interaction \cite{franosch2011resonances}. Here, the presence of the viscous fluid allows for a viable path to formulate the Brownian dynamics of the nanomechanical resonator consistently~\cite{paul2004stochastic,paul2006stochastic,clark2010spectral}: first, the dissipation of the resonator is found from fluid dynamics; then, the fluctuation-dissipation theorem  is used for the calculation of the PSD of the resonator fluctuations. Since the fluidic dissipation is  frequency dependent, the force noise PSD is also ``colored," and the resonator fluctuations deviate substantially from the first-order approximation discussed above.   In nearly all work so far, the dissipation in viscous fluids has been  assumed to be mode independent and spatially uniform, resulting in a Brownian force noise that is spatially uncorrelated. This assumption of spatial homogeneity again leads to formulas expressible as a sum in terms of the individual eigenmodes.  There are notable papers, where the experimental noise data have been successfully fitted with colored PSDs. However, these experiments typically do not extend beyond the first  mode of the elastic structure~\cite{clarke2006stochastic, bellon2008thermal, pierro2019thermal, ari2020nanomechanical} and are thus not very insightful on the spatial nature of the force noise. 

The topic of this manuscript is the  Brownian dynamics of a nanomechanical resonator in a viscous fluid. In particular, we investigate  how  the Brownian force driving the nanomechanical resonator is correlated in time and space in a viscous fluid.  To this end, we derive an expression for the PSD of the displacement fluctuations of an elastic nanomechanical beam under tension in a viscous fluid, assuming a frequency-dependent but spatially homogeneous viscous dissipation. This results in a PSD that is the sum of the PSDs of the uncorrelated fluctuations of individual   eigenmodes. We validate this theory by experiments performed on nanomechanical beams under tension immersed in air and water.  Using solely experimental parameters, we obtain excellent agreement between the experimental data and theory. This agreement, up to the twelfth eigenmode in air and seventh eigenmode in water, validates our overarching assumptions: the Brownian force noise has a colored PSD due to the memory of the fluid but can be approximated to be uncorrelated in space. 

\begin{figure}
    \includegraphics[width=3.375in]{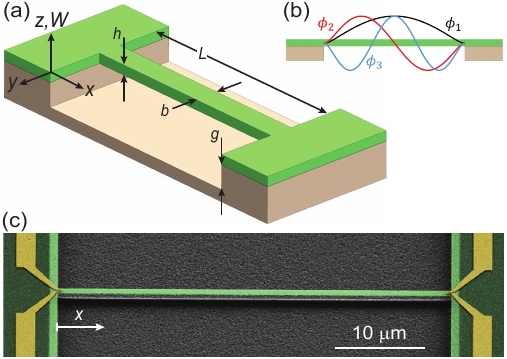}
    \caption{(a) A doubly-clamped beam with length $L$, width $b$, thickness $h$, and gap $g$ between the beam and substrate. (b) Illustration of the first three eigenfunctions $\phi_n(x)$. (c) SEM image of a doubly-clamped silicon nitride beam under tension with $L\times b\times h \!=\! 50~\rm\mu m\times900~nm\times93~nm$ and $g \!=\! 2~\rm\mu m$. The silicon nitride layer is shown in green, with the suspended part in light green. The silicon substrate and the gold layer are  gray and yellow, respectively.}    
    \label{Figure_1}
\end{figure}

\section{Theory}

We start with the  equation of motion for a beam under tension  driven by a deterministic external force in a viscous fluid. The respective linear dimensions of the beam are $L \times b \times h$ along the $xyz$ axes [Fig.~\ref{Figure_1}(a)], and $\mu=\rho_s b h$ is the mass per unit length with $\rho_s$ being the density. The flexural displacement of the beam, $W(x,t)$, along the $z$ axis at position $x$ and time $t$ is given by
\begin{multline}
\frac{EI}{L^4}\frac{\partial^4 W(x,t)}{\partial x^4} - \frac{F_T}{L^2} \frac{\partial^2 W(x,t)}{\partial x^2} + \mu \frac{\partial^2 W(x,t)}{\partial t^2} =  \\ F_f(x,t) + F_d(x,t).
\label{eq:euler-bernoulli}
\end{multline}
The     $x$ coordinate has been normalized by $L$ such that $0 \!\le x \! \le 1$. In Eq.~(\ref{eq:euler-bernoulli}), $E$ is the Young's modulus, $I$ is the area moment of inertia, $F_f(x,t)$ is the force per unit length of the fluid acting on the beam, $F_d(x,t)$ is the external drive force per unit length, and $F_T$ is the tension~\cite{ti2021frequency}. The beam has fixed boundaries such that $W(0,t)\!=\!W(1,t)\!=\! W'(0,t)\!=\! W'(1,t)=\! 0$, where a prime indicates an $x$ derivative.

It will be useful to proceed in the frequency domain~\cite{barbish2022dynamics} using the Fourier transform pair 
\begin{eqnarray}
\hat{W}(x,\omega) &=& \int_{-\infty}^{\infty} W(x,t) e^{i \omega t} dt, \\
W(x,t) &=& \frac{1}{2 \pi} \int_{-\infty}^{\infty} \hat{W}(x,\omega) e^{-i \omega t} d \omega,
\end{eqnarray}
where $\omega$ is the angular frequency. This leads to the transformed differential equation  
\begin{multline}
\frac{EI}{L^4}\frac{\partial^4 \hat{W}(x,\omega)}{\partial x^4} - \frac{F_T}{L^2} \frac{\partial^2 \hat{W}(x,\omega)}{\partial x^2}\\
- \omega^2 \left[ \mu + \frac{\pi}{4} \rho_f b^2 \Gamma(\omega) \right] \hat{W}(x,\omega) = \hat{F}_d(x,\omega)
\label{eq:eom-fourier}
\end{multline} 
for the Fourier component  $\hat{W}(x,\omega)$  at $\omega$. In Eq.~(\ref{eq:eom-fourier}), we have described the force due to the fluid as
\begin{equation}
\hat{F}_f(x,\omega) = \frac{\pi}{4} \rho_f \omega^2 b^2 \Gamma(\omega) \hat{W}(x,\omega),
\label{eq:hydro-force}
\end{equation}
where $\rho_f$ is the fluid density and $\Gamma(\omega)$ is  the complex hydrodynamic function for a blade,  derived from Stokes' oscillating cylinder theory ~\cite{schlichting1932berechnung,retsina1987theory,sader1998frequency}. In Eq.~(\ref{eq:hydro-force}), $\Gamma(\omega)$  quantifies the mass loading and viscous damping of the fluid acting on the beams. To obtain $\Gamma(\omega)$, one starts with the complex hydrodynamic function for an  infinitely-long oscillating cylinder,
\begin{equation}
    \Gamma_c\left(\rm{Re}_\omega\right) = 1 + \frac{4iK_1\left(-i\sqrt{i\rm{Re}_\omega}\right)}{\sqrt{i\rm{Re}_\omega}K_0\left(-i\sqrt{i\rm{Re}_\omega}\right)},
\label{eq:Gamma_c}
\end{equation}
where $K_0$ and $K_1$ are respectively the zeroth and first order modified Bessel functions of the second kind \cite{rosenhead1963laminar}. The argument of $\Gamma_c$ is the frequency-dependent Reynolds number, ${\rm{Re}}_\omega=\frac{\rho_f\omega b^2}{4\eta_f}$, where $\eta_f$ is the dynamic viscosity of the fluid. To account for the rectangular cross-section of the beams,  one then applies a small frequency-dependent correction factor  to $\Gamma_c$ \cite{sader1998frequency}.  It can be deduced from Eq.~(\ref{eq:Gamma_c}) that the only parameters in $\Gamma(\omega)$ are $b$ and  $\eta_f$, which are both constants. Thus, $\Gamma(\omega)$ is assumed to be independent of position $x$ as well as the mode-shape of the beam.

We solve Eq.~(\ref{eq:eom-fourier}) using the eigenfunction expansion~\cite{sader1998frequency}, 
\begin{equation}
\hat{W}(x,\omega) = \sum_{n=1}^{\infty} f_n(\omega) \phi_n(x),
\end{equation}
where $n$ is the mode number, $f_n(\omega)$ describes the frequency dependence, and $\phi_n(x)$ are the orthonormal eigenfunctions of the beam with tension [Fig.~\ref{Figure_1}(b)]. Expressions for  $\phi_n(x)$ and the eigen-frequencies $\omega_n/2 \pi$ for a doubly clamped beam with tension are available \cite{barbish2022dynamics, bokaian1990natural,stachiv2014impact,ari2020nanomechanical}. We note that both $\phi_n(x)$ and $\omega_n/2 \pi$ are found from the dissipationless equation of motion. The influence of the tension force on $\phi_n(x)$ and $\omega_n$ can be quantified in terms of the nondimensional tension parameter $U$, where $U = \frac{F_T}{2EI/L^2}$. The dynamics respectively becomes that of an Euler-Bernoulli beam and a string for $U\to 0$ and   $U\gg 1$.

Using the orthogonality of $\phi_n(x)$, the solution to Eq.~(\ref{eq:eom-fourier}) can be expressed as 
\begin{equation}
\hat{W}(x,\omega) = \frac{L^4} {EI} \sum_{n=1}^{\infty} \frac{ \int_0^1 \hat{F}_d(x',\omega) \phi_n(x')  dx'} {\Omega_n^2 - B^4(\omega)} \phi_n(x).
\end{equation}
$\Omega_n$ are the nondimensional eigen-frequencies defined as
\begin{equation}
\Omega_n \!=\! \frac{\omega_n}{\alpha/L^2},     
\label{eqn:nondim-eigenfrequency}
\end{equation}
where $\alpha\!=\!(EI/\mu)^{1/2}$. The complex function $B(\omega)$ contains the dissipation and added mass, and is given by
\begin{equation}
B^4(\omega) = \Omega_1^2 \left( \frac{\omega}{\omega_1} \right)^2 \left[ 1 + T_0 \Gamma(\omega) \right],
\end{equation}
where $\omega_1$ is the fundamental eigen-frequency of the beam in the absence of  fluid, i.e., dissipation and added mass. The mass loading parameter, $T_0 \!=\! \frac{\pi}{4} \frac{ \rho_f b}{\rho_s h}$, is the ratio of the mass of a cylinder of fluid with diameter $b$ to the mass of the beam.

In order to connect with the fluctuation-dissipation theorem, we next calculate the susceptibility, ${\chi}(x_0,t)$, which we define as the time-dependent displacement of the beam measured at position $x_0$ due to the application of a unit impulse of force  at the same position $x_0$. Thus, we specify 
\begin{equation}
F_d(x,t) = \frac{1}{L} \delta(x-x_0)\delta(t),
\end{equation}
which becomes 
\begin{equation}
\hat{F}_d(x,\omega) = \frac{1}{L} \delta(x-x_0)
\end{equation}
in the frequency domain, with $\delta$ being the Dirac delta function. Hence, $\hat{\chi}(x_0,\omega) = \hat{W}(x_0,\omega)$, and we obtain
\begin{equation}
\hat{\chi}(x_0,\omega) = \frac{L^3} {EI} \sum_{n=1}^{\infty} \frac{ \int_0^1 \delta(x'-x_0) \phi_n(x')  dx'} {  \Omega_n^2 - B^4(\omega) } \phi_n(x_0),
\end{equation}
which can be expressed as 
\begin{equation}
\hat{\chi}(x_0,\omega) = \frac{L^3} {EI} \sum_{n=1}^{\infty} \frac{ \phi_n^2(x_0)} {\Omega_n^2 - B^4(\omega)}.
\end{equation}
Using $\omega_n/\omega_1 = \Omega_n/\Omega_1$, defining $\tilde{\omega}_n = \omega/\omega_n$, and simplifying further yields
\begin{multline}
\hat{\chi}(x_0,\omega) = \sum_{n=1}^{\infty} \frac{1} {k_n(x_0)} \\ \times \frac{1} {1 - \tilde{\omega}_n^2 (1+ T_0 \Gamma'(\omega)) -  i \tilde{\omega}_n^2 T_0 \Gamma''(\omega)},
\label{eq:chi-hat}
\end{multline}
where $\Gamma'(\omega)$ and $\Gamma''(\omega)$ are the real and imaginary parts of $\Gamma(\omega)$, respectively. The effective spring constant of mode $n$, when measured at $x_0$, is represented as $k_n(x_0)$; $k_n(x_0)$ can be consistently determined by ensuring that the kinetic energy of the spatially extended oscillating beam with mode shape $\phi_n(x)$ equals that of a lumped system measured at $x_0$. This yields
\begin{equation}
k_n(x_0) = \frac{m \omega_n^2}{\phi_n^2(x_0)},
\label{eq:k_n}
\end{equation}
where $m=\mu L$ is the nominal mass of the beam. 

The PSD of the Brownian fluctuations of the beam at axial position $x_0$ can be directly expressed using the fluctuation-dissipation theorem~\cite{callen1951irreversibility,callen1952theorem} as
\begin{equation}
G_W(x_0,\omega) = \frac{4 k_B T}{\omega} \hat{\chi}''(x_0,\omega),
\label{eq:noise}
\end{equation}
where $\hat{\chi}''(x_0,\omega)$ is the imaginary part of $\hat{\chi}(x_0,\omega)$, $k_B$ is Boltzmann's constant, and $T$ is the temperature. The subscript $W$ on $G_W$ indicates that this is the spectral density of the fluctuations in the flexural displacement $W(x,t)$ of the beam along the $z$ axis.

Taking the imaginary part of Eq.~(\ref{eq:chi-hat}) to find $\hat{\chi}''(x_0,\omega)$ and inserting this into Eq.~(\ref{eq:noise}), we obtain the desired result as
\begin{multline}
      G_W(x_0, \omega) = 4 k_B T \sum_{n=1}^{\infty} \frac{1}{k_n(x_0) \omega_n} \\ \times \frac{\tilde{\omega}_n  T_0 \Gamma''(\omega)}{\left[1 - \tilde{\omega}_n^2 (1 + T_0 \Gamma'(\omega)) \right]^2 + \left[  \tilde{\omega}_n^2  T_0 \Gamma''(\omega)\right]^2}. 
\label{eq:gx} 
\end{multline}
This expression yields the total PSD for the displacement fluctuations of the beam at frequency $\omega$ and position $x_0$. We emphasize that $G_W(x_0, \omega)$ in Eq.~(\ref{eq:gx}) is obtained as a sum over individual eigenmodes. This is because $\Gamma(\omega)$ is assumed to be spatially homogeneous and independent of mode number $n$~\cite{van2007frequency}.

\begin{figure*}
    \centering
    \includegraphics[width=6.58in]{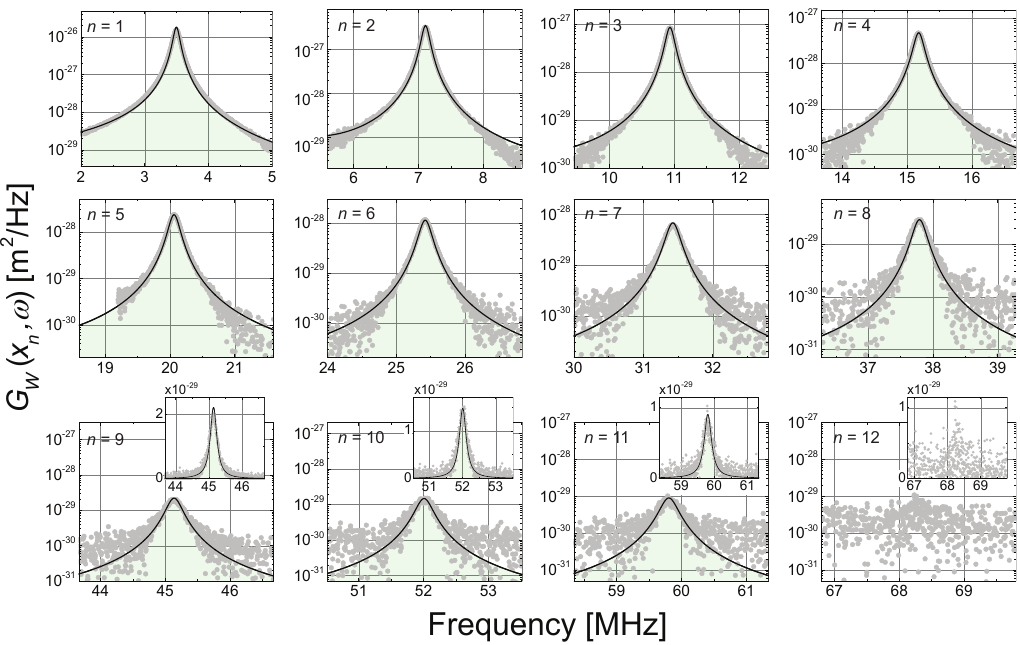}
    \caption{PSDs of the displacement fluctuations $G_W(x_n,\omega)$ plotted using semi-logarithmic axes for the first twelve modes of a beam with $L=30~\rm \mu m$ in air. The measurement is performed at $x\!=\!x_n$, i.e., an antinode of mode $n$. The continuous lines are fits based on Eq.~(\ref{eq:gx}). The insets for $n \!=\! 9$ to $12$ show $G_W(x_n,\omega)$ using linear axes. Because the peak of the twelfth mode is barely resolved, we do not determine values for $\omega_{12}/2\pi$, $k_{12}(x_{12})$, and $Q_{12}^{(a)}$.}   
    \label{Figure_2}
\end{figure*}

\section{Experiments}

Our experiments are performed on  silicon nitride doubly-clamped beams of  $b\!=\!900~\rm nm$,  $h\!=\!93~\rm nm$, and three different lengths of $L\!=\!30,~40$ and $50~{\rm \mu m}$; there is a gap of $g\!=\!2~\rm\mu m$ between each beam and the substrate. Fig.~\ref{Figure_1}(c) shows a scanning electron microscope (SEM) image of a beam with $L\!=\!50~\rm\mu m$. All the beams are from the same fabrication batch. The beams are under tension as inferred from their resonance frequencies in vacuum~\cite{ti2021frequency}. We determine $\mu\!=\!\rho_sbh \!=\! 2.66 \!\times\! 10^{-10} ~ \rm kg/m$, with the density measured as $\rho_s \!=\! 2960~\rm kg/m^3$~\cite{ti2021frequency}. The beams also have u-shaped gold nanoresistor patterns near their anchors for other experiments~\cite{ari2020nanomechanical,ti2021frequency}.

We measure the displacement fluctuations of the beams using a path-stabilized homodyne Michelson interferometer. The diffraction limited HeNe laser spot with a diameter of  $\sim650\pm10$ nm (FWHM) is positioned on the beam using an \textit{XYZ} precision stage. The typical powers incident on the beam and the photodetector are $\sim600~\rm\mu W$ and $\sim 1~\rm m W$, respectively, with a shot noise limited displacement sensitivity of $\sim5~\rm fm/\sqrt{Hz}$. We calibrate the system against the wavelength of the laser \cite{wagner1990optical} and operate at the point of optimal sensitivity.  For each measurement taken at a given position $x_0$ on a beam, a second measurement is taken with the same parameters near the anchor of the beam to determine the background noise level in our measurements, e.g., due to the low-frequency laser noise or cable resonances. We assume that the beam fluctuations and the background noise are uncorrelated, and  subtract this background from the noise PSD measured on the beam \cite{kara2015nanofluidics}. This allows us to resolve the beam fluctuations down to $\sim2~\rm fm/\sqrt{Hz}$. 

\begin{figure*}
    \includegraphics[width=6.58in]{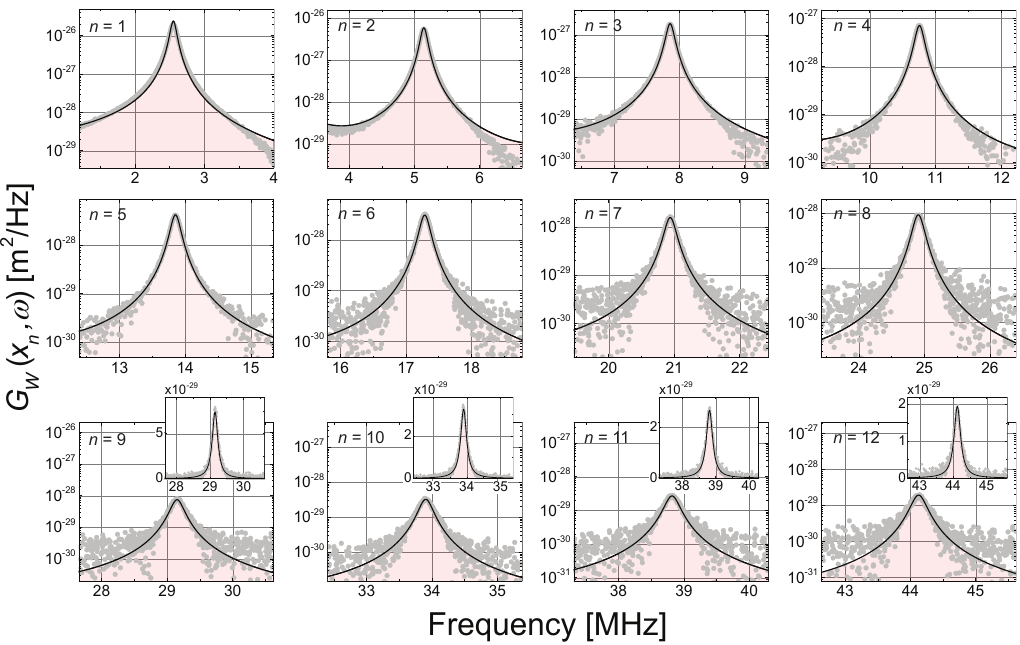}
    \caption{PSDs of the displacement fluctuations $G_W(x_n,\omega)$ plotted using semi-logarithmic axes for the first twelve modes of a beam with  $L=40~\rm \mu m$ in air. The measurement is performed at $x\!=\!x_n$, i.e., an antinode of mode $n$. The continuous lines are fits based on Eq.~(\ref{eq:gx}). The insets for $n \!=\! 9$ to $12$ show $G_W(x_n,\omega)$ using linear axes.}    
    \label{Figure_3}
\end{figure*}

\begin{figure*}
    \centering
    \includegraphics[width=6.58in]{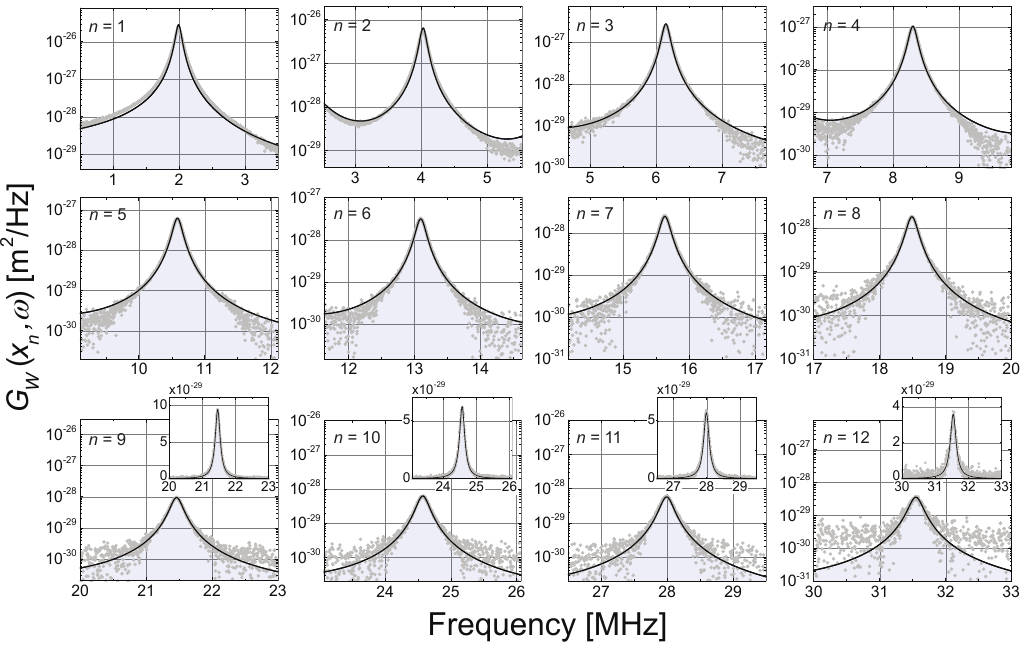}
    \caption{PSDs of the displacement fluctuations $G_W(x_n,\omega)$ plotted using semi-logarithmic axes for the first twelve modes of a beam with $L=50~\rm \mu m$ in air. The measurement is performed at $x\!=\!x_n$, i.e., an antinode of mode $n$. The continuous lines are fits based on Eq.~(\ref{eq:gx}). The insets for $n \!=\! 9$ to $12$ show $G_W(x_n,\omega)$ using linear axes.}   
    \label{Figure_4}
\end{figure*}

The finite size of the optical spot introduces errors into the measurements \cite{kouh2005diffraction}. A significant source of error is  the curvature of the beam, especially in higher modes. By computing the overlap of the Gaussian optical spot with the beam mode,  we estimate this error to be less than $10\%$  for the twelfth mode of our $30$-$\rm\mu m$-long beam, which has the largest curvature in all our experiments. In  water measurements, it also becomes problematic to  position the optical spot precisely at the desired locations on the beam.  To find $x_0=0.50$ (center) and $x_0=0.25$ ($L/4$) positions on the beam, we maximize the $n=1$ and $n=2$ peaks, respectively. 

\section{Results}

\subsection{Vacuum}

We first measure the resonance frequencies of the eigenmodes of the beams in vacuum ($p<10^{-6}~\rm bar$), as listed in Table~\ref{Table_S1}.  These peak frequencies in vacuum, $\omega_n^{(v)}/2\pi$, should be very close to the eigen-frequencies of the dissipationless beam,  $\omega_n/2\pi$. We also extract quality factors in vacuum from Lorentzian fits and find that all $Q_n^{(v)}\gtrsim 10^3$. 

\begin{figure}
    \includegraphics[width=3.37in]{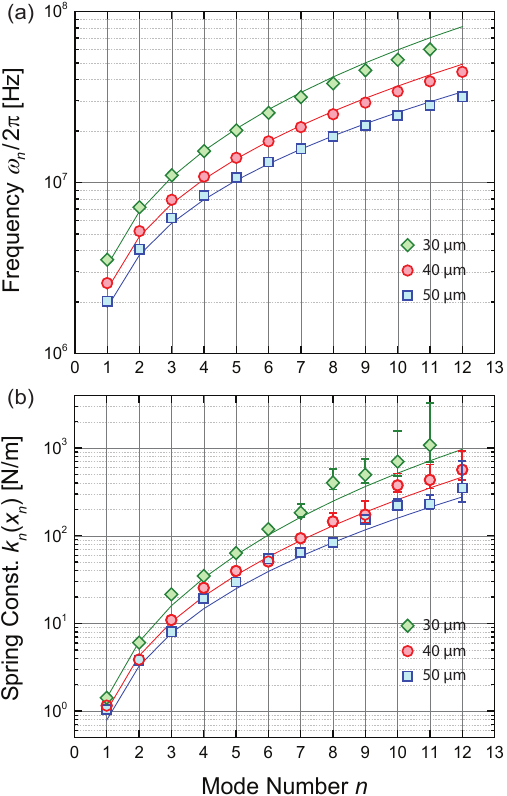}
    \caption{(a) Eigen-frequencies $\omega_n/2\pi$ and (b) effective spring constants $k_n(x_n)$  for   beams with $L=30~\rm \mu m$, $L=40~\rm \mu m$, and $L=50~\rm \mu m$.  Experimental values and theoretical predictions are shown by symbols and continuous lines, respectively. Typical error bars for $k_n(x_n)$ are smaller than the  symbols unless shown explicitly.}    
    \label{Figure_5}
\end{figure}

\begin{figure*}
    \includegraphics[width=6.58in]{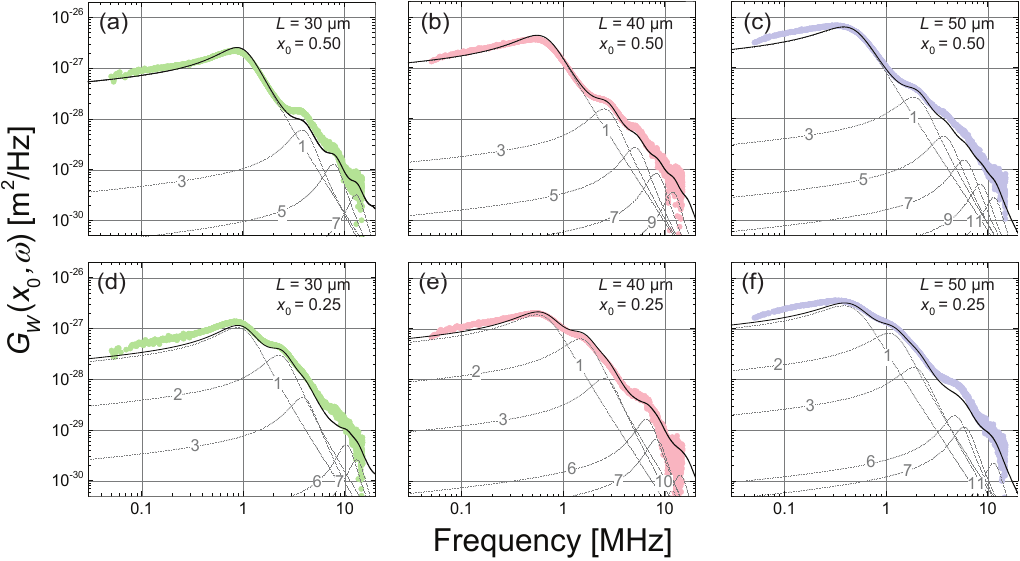}
    \caption{PSDs of the displacement fluctuations $G_W(x_0,\omega)$ at $x_0=0.50$ (a-c) and $x_0=0.25$ (d-f) for three beams with different lengths in water. The continuous lines are predictions of Eq.~(\ref{eq:gx}) based on experimental parameters. The dotted lines show the  contributions of individual modes $n$ to $G_W(x_0,\omega)$. Modes not shown do not contribute significantly at the particular $x_0$.} 
    \label{Figure_6}
\end{figure*}

\subsection{Air}

Next, we  examine the displacement fluctuations in air. We first identify the frequency at which our system transitions from viscous to molecular flow. For this system, the transition frequency $\omega_c/2\pi$ can be found using  ${\omega_c \tau}+{\lambda \over b}\approx1$, where  $\tau$ is the relaxation time and $\lambda$ is the mean free path in the fluid \cite{kara2015nanofluidics,kara2017generalized}. In air, $\tau\approx1~\rm ns$  and $\lambda\approx68~\rm nm$ \cite{kara2015nanofluidics,kara2017generalized}. The relevant length scale, $b= 900~\rm nm$, is the same for all our devices. We thus find $\omega_c/2\pi\approx100~\rm MHz$ in air. Therefore,  our measurements are mostly within the viscous regime. 

Figs.~\ref{Figure_2}, \ref{Figure_3}, and \ref{Figure_4} respectively show the PSDs of the first twelve modes of beams with  $L=30, 40$ and $50$ $\rm\mu m$. The low dissipation in air results in distinctly separated peaks in the PSDs. Each PSD is measured at an antinode ($x=x_n$) of mode $n$ as a function of frequency near the peak frequency, $\omega_n^{(a)}/ 2\pi$, in air. The relatively large modal quality factors, $17 \le Q_n^{(a)}\le 200$, allow us to determine the eigen-frequencies $\omega_n/2\pi$, the effective spring constants $k_n(x_n)$, and the theoretical $G_W(x_{n},\omega)$ curves in a self-consistent manner. To this end, we first use the equipartition of energy to find $k_n(x_{n})$ from $\frac{1}{2}k_n(x_{n}) \left<W_n^2 (x_{n})\right> \!=\! \frac{1}{2} k_B T$, where the mean-squared fluctuation amplitude, $\left<W_n^2 (x_{n})\right>$, is the numerical integral of the experimental $G_W(x_{n},\omega)$ data over frequency. To find the theoretical $G_W(x_{n},\omega)$ curve, we calculate $T_0$ and $\Gamma(\omega)$ using the density and viscosity of air at room temperature. We then insert the experimental $k_n(x_{n})$ values along with $T_0$ and $\Gamma(\omega)$  into Eq.~(\ref{eq:gx}) and calculate $G_W(x_{n},\omega)$, treating $\omega_n$ as a fit parameter. The best fits are shown as continuous lines in Figs.~\ref{Figure_2}, \ref{Figure_3}, and \ref{Figure_4}.

We re-emphasize that, except for $\omega_n$, the fits (continuous lines) in Figs.~\ref{Figure_2}, \ref{Figure_3}, and \ref{Figure_4} are solely determined using known or measured quantities of the beam and the surrounding fluid.   The values of $\omega_n/2\pi$  found by fitting are typically slightly higher than  $\omega_n^{(a)}/ 2\pi$ (Table~\ref{Table_S1}). This is expected because $\omega_n/2\pi$ are the eigen-frequencies without any fluid loading.  Our vacuum measurements support this observation (Table~\ref{Table_S1}): the vacuum frequencies $\omega_n^{(v)}/2\pi$ are all slightly larger than $\omega_n^{(a)}/ 2\pi$. There are very small discrepancies (typically $<1.5\%$) between the eigen-frequencies determined by fitting, i.e., $\omega_n/2\pi$, and those obtained from vacuum measurements, $\omega_n^{(v)}/2\pi$.  We attribute these small discrepancies to the fact that the eigen-frequencies of nanomechanical resonators are easily perturbed by external factors, e.g., the accumulation of adsorbates or changes in the temperature and humidity of the environment. Since the resonance  is very sharply peaked in air ($Q_n^{(a)}>10$) and in vacuum ($Q_n^{(v)}\sim10^3$), these perturbations result in small but noticeable frequency shifts from measurement to measurement.

\begin{table*}
\vspace{10pt}
\rotatebox{90}{
\begin{minipage}{0.95\textheight}
\caption{Eigen-frequencies $\omega_n/2\pi$ are from fitting our air measurements to our model; peak frequencies, $\omega_n^{(v)}/2\pi$ and $\omega_n^{(a)}/2\pi$, are  measured in vacuum and  air, respectively. The effective stiffness $k_n(x_n)$ is found at an antinode, and quality factors $Q_n^{(a)}$ in air are found from Lorentzian fits. All frequency values are in MHz. The modal peak for $n=12$ is barely resolvable in the PSD [Fig.~\ref{Figure_2}].} \label{Table_S1}
\renewcommand{\arraystretch}{2}
\newcolumntype{Y}{>{\centering\arraybackslash}X}
\begin{tabularx}{\textwidth}{Y|YYYYY|YYYYY|YYYYY}
Mode & \multicolumn{5}{c|}{$L=30~\rm \mu m$} & \multicolumn{5}{c|}{$L=40~\rm \mu m$} & \multicolumn{5}{c}{$L=50~\rm \mu m$}\\[5pt]
 & $\omega_n\over 2\pi$ & $\omega_n^{(v)}\over 2\pi$ & $\omega_n^{(a)}\over 2\pi$ & $k_n(x_n)$ & $Q_n^{(a)}$ & $\omega_n\over 2\pi$ & $\omega_n^{(v)}\over 2\pi$ & $\omega_n^{(a)}\over 2\pi$ & $k_n(x_n)$ & $Q^{(a)}$ & $\omega_n\over 2\pi$ & $\omega_n^{(v)}\over 2\pi$ & $\omega_n^{(a)}\over 2\pi$ & $k_n(x_n)$ & $Q^{(a)}$\\[10pt]
 \hline
1 & 3.531 & 3.538 & 3.501 & 1.42 & 28 & 2.577 & 2.596 & 2.553 & 1.16 & 22 & 2.017 & 2.019 & 1.995 & 1.04 & 17\\
2 & 7.158 & 7.224 & 7.111 & 6.05 & 50 & 5.182 & 5.310 & 5.145 & 3.86 & 40 & 4.065 & 4.073 & 4.030 & 3.81 & 35\\
3 & 10.998 & 11.009 & 10.934 & 21.42 & 69 & 7.906 & 7.997 & 7.856 & 10.94 & 56 & 6.194 & 6.168 & 6.151 & 7.99 & 46\\
4 & 15.260 & 15.293 & 15.180 & 34.65 & 86 & 10.817 & 10.968 & 10.755 & 25.46 & 70 & 8.356 & 8.326 & 8.303 & 19.11 & 58\\
5 & 20.159 & 20.141 & 20.067 & 63.28 & 104 & 13.924 & 14.073 & 13.851 & 39.62 & 83 & 10.649 & 10.598 & 10.585 & 29.64 & 68\\
6 & 25.532 & 25.605 & 25.420 & 118.67 & 121 & 17.373 & 17.537 & 17.289 & 50.77 & 98 & 13.166 & 13.227 & 13.094 & 54.49 & 81\\
7 & 31.556 & 31.461 & 31.423 & 183.02 & 125 & 21.037 & 21.121 & 20.938 & 93.52 & 110 & 15.717 & 15.725 & 15.636 & 64.15 & 92\\
8 & 37.940 & 37.784 & 37.789 & 405.19 & 141 & 25.017 & 25.175 & 24.910 & 145.15 & 120 & 18.579 & 18.598 & 18.487 & 83.14 & 102\\
9 & 45.306 & 45.086 & 45.132 & 495.15 & 150 & 29.275 & 29.354 & 29.156 & 173.99 & 131 & 21.562 & 21.564 & 21.462 & 153.41 & 112\\
10 & 52.195 & 52.392 & 52.004 & 699.80 & 155 & 34.043 & 34.265 & 33.904 & 375.16 & 137 & 24.682 & 24.695 & 24.573 & 220.33 & 119\\
11 & 60.014 & 59.904 & 59.802 & 1077.36 & 167 & 38.970 & 38.879 & 38.817 & 375.16 & 144 & 28.112 & 27.988 & 27.991 & 220.33 & 128\\
12 & -- & 68.342 & 68.293 & -- & -- & 44.291 & 44.234 & 44.120 & 564.47 & 153 & 31.678 & 31.565 & 31.548 & 349.46 & 136\\
\end{tabularx}
\end{minipage}}
\end{table*}

\subsection{ Eigen-Frequencies and Spring Constants}

We next compare the experimental values for  $\omega_n/2 \pi$ and $k_n(x_n)$ with theoretical predictions of Euler-Bernoulli beam theory with tension~\cite{barbish2022dynamics}. We estimate the magnitude of the tension force $F_T$ by comparing the eigen-frequencies obtained from experiments, $\omega_n/2\pi$, with those from theory, $\omega_n^{(t)}/2\pi$. To determine $\omega_n^{(t)}/2\pi$, we turn to the characteristic equation, which relates $\omega_n^{(t)}/2\pi$ to the unknown tension \cite{bokaian1990natural} in terms of the non-dimensional parameters $\Omega_n$ and $U$. In our calculations, we use nominal beam dimensions as well as $E$ and $\rho_s$ values for SiN. Since reported values for $E$ have a  large uncertainty, $200~ {\rm GPa}\lesssim E \lesssim380~\rm GPa$ \cite{tai1990measurement, zhang2000microbridge,kuhn2000fracture}, we take the Young's modulus as $E=300~\rm GPa$. The density has been measured as $\rho_s=2960\pm30~\rm kg/m^3$ \cite{ti2021frequency}. For each beam, we sweep the value of $F_T$, solve for  $\omega_n^{(t)}/2\pi$ for each $F_T$, and then compute the  error 
\begin{equation}
    \varepsilon=\sum_{n=1}\frac{|\omega_n-\omega_n^{(t)}|^2}{|\omega_n|^2}
\label{eq:error}
\end{equation}
between $\omega_n$ and $\omega_n^{(t)}$, where $n$ encompasses the first twelve modes. The minimum  $\varepsilon$ provides the experimental value of $F_T$. Because all beams are on the same chip, we average the values of $F_T$ for each of our three beams. We thus find $F_T=7.43~\rm \mu N$.  In Fig.~\ref{Figure_5}(a), we show experimental $\omega_n/2 \pi$ data (symbols) and theoretical predictions  (continuous lines) using $F_T=7.43~\rm \mu N$   on semi-logarithmic axes for all three beams. 

To find the theoretical $k_n(x_n)$ at an antinode, we use Eq.~(\ref{eq:k_n}). To this end, we  calculate $\phi_n(x)$ for each beam~\cite{barbish2022dynamics} using $F_T$ and $\omega_n/ 2\pi$; we use nominal $m$, measured $\omega_n$, and calculated $\phi_n(x_n)$ to determine $k_n(x_n)$. We show the experimental and theoretical values for $k_n(x_n)$ in Fig.~\ref{Figure_5}(b). The experimental spring constants match predictions closely over two orders of magnitude  for $n\lesssim 7$.  With our knowledge of $\phi_n(x)$ and $k_n(x_n)$, we can  determine  $k_n(x_0)$ for any  position $x_0$ along the beam.

The differences between experimental and theoretical values of $\omega_n$ and $k_n(x_n)$ are most likely due to imperfections, such as the presence of the gold layer and the undercuts beneath the anchors.  There is  also an estimated  10\% error  in $F_T$ due to the fact that we do not know the exact value of $E$ \cite{ari2020nanomechanical}. Thus, it is more justifiable to use the $\omega_n$ and $k_n(x_n)$  values directly obtained from experiments rather than those calculated from elasticity theory.

\subsection{Water}

Our beams are then immersed in water, where we measure the PSDs of displacement fluctuations, $G_W(x_0,\omega)$, over the continuous frequency range of $50 ~{\rm kHz} $ to $15~\rm MHz$ at two positions on the beam,  $x_0=0.25$ and $x_0=0.50$. Fig.~\ref{Figure_6} shows $G_W(x_0,\omega)$ as a function of frequency at these two positions for all three beams. The first position, $x_0=0.50$, corresponds to an antinode of all odd modes and a node of all even modes [Fig.~\ref{Figure_6}(a-c)]; the second position, $x_0=0.25$, is close to the antinode of the second mode [Fig.~\ref{Figure_6}(d-f)]. Low quality factors ($Q\gtrsim1$) in water result in broad and overlapping peaks; the peak frequencies are significantly lower than those in air.

Next, we compare our measurements in water to our theoretical expression for $G_W(x_0,\omega)$. The continuous curves in Fig.~\ref{Figure_6} show predictions based on Eq.~(\ref{eq:gx}). Here, we  use the $\omega_n$ values found above and determine $k_n(x_0)$ from $k_n(x_n)$ after correcting for the position dependence via Eq.~(\ref{eq:k_n}). We then compute $T_0$ and $\Gamma(\omega)$ using the density and viscosity of water, and combine all factors in Eq.~(\ref{eq:gx}) to generate the curves. The dotted curves show the PSDs of individual modes; the continuous curve is the sum of the first $12$ modes. A strong agreement between experiment and theory is evident for $n\lesssim7$; for $n\gtrsim 7$, the beam fluctuations remain below our resolution limit. The theory predicts the peak frequencies and the noise power levels  accurately.

The positioning error mentioned in the third paragraph in Section~III affects both $k_n(x_0)$ and the measured $G_W(x_0,\omega)$. This error becomes more pronounced at higher frequencies. In principle, the theory curves can be further improved by treating the measurement position as a fit parameter.

\section{Discussion}

Eq.~(\ref{eq:gx}) describes the Brownian dynamics of a nanomechanical beam  in a viscous fluid and indicates that, at a given frequency, the total noise is found by adding the noise PSDs in different eigenmodes.   The underlying assumption is that the Brownian force noise is delta-function correlated in space~\cite{newland2012introduction,sader1998frequency}. The form of Eq.~(\ref{eq:gx}), i.e., the summation over uncorrelated eigenmodes,  should remain unchanged for any mechanical system as long as  the  dissipation is uniform in space. For our system, the dissipation in  Eq.~(\ref{eq:hydro-force})  from the cylinder model is indeed spatially homogeneous. The remarkable agreement between experiment and theory in Fig.~\ref{Figure_6} for the first seven modes suggests that the  cylinder model remains accurate---to within our experimental resolution. In other words,  the  frequency dependence and the spatial homogeneity of the dissipation in the model are both validated by our experiments. 

Hydrodynamic fluctuations in a simple fluid are typically assumed to be delta-function correlated in space~\cite{landau1980course}. However, the situation is different for the nanomechanical beam immersed in a fluid: the flow around the structure and the fluid-structure interactions are expected to result in spatial correlations in the force noise, eventually leading to observable deviations from Eq.~(\ref{eq:gx}) for higher modes. This expectation is consistent with the fact that the viscous dissipation of the oscillating cylinder model is not accurate for higher modes. As the flow in the axial direction becomes more appreciable with increasing mode number, the dissipation becomes mode dependent and non-uniform~\cite{liem2021nanoflows, van2007frequency}.  However, the agreement between predictions and experiments suggests that the axial flow is negligible in our parameter space. The smallest length scale probed in our beams is comparable to the smallest resolved modal wavelength of $\sim9~\rm\mu m$.

$G_W(x_0,\omega)$ should also be affected by the presence of the nearby substrate due to the squeeze flow between the beam and the substrate. In our theory, we have neglected the presence of the substrate. This can be corrected using numerical simulations~\cite{liem2021nanoflows}: we estimate a decrease in $\Gamma'(\omega/2\pi=0.1~\rm MHz)$ by $\sim30\%$ and an increase in $\Gamma''(\omega/2\pi \!=\! 0.1~\rm MHz)$ by $\sim30\%$, which leads to an increase in $G_W(\omega/2\pi \!=\! 0.1~\mathrm{MHz},x_0)$ of  $\lesssim30\%$ depending on  $L$ of the beam. This additional damping should further broaden the fundamental mode and induce an additional redshift of the peak~\cite{green2005frequency,clarke2006stochastic}, which we observe in measurements. At higher frequencies, as the viscous boundary layer thickness $\delta$ becomes $\delta/g \!\ll\! 1$, this effect disappears~\cite{clarke2005drag,pierro2019thermal}. 

To experimentally observe the noise correlation effects due to fluid-structure interaction, it would be necessary to resolve the fluctuations in higher modes. To this end, one should first determine the viscous dissipation from fluid dynamics and asses the regime where the  dissipation becomes nonhomogenous, e.g., due to axial flows.  Unfortunately, analytical solutions cannot be found for most experimental geometries, making numerical simulations necessary. Once the flow regimes around the structure are determined,  one could design and fabricate   structures with softer spring constants $k_n$ such that the Brownian motions in higher modes can be resolved.

Nonlinearities in the elastic potential can modify the Brownian dynamics of the nanomechanical beam \cite{dykman1971classical, gieseler2013thermal}. Achieving the nonlinear limit for  thermal fluctuations of a mode requires  a very high modal $Q_n$ and  a  very soft modal spring \cite{gieseler2013thermal}.  We estimate that the very low $Q_n$  in fluids along with the large $k_n$ of the beams makes nonlinear  effects negligible in our experiments.   As an order of magnitude comparison,   the fundamental mode of our 50-$\rm\mu m$-long beam  displays nonlinear behavior at amplitudes $\gtrsim 20~\rm nm$ \cite{postma2005dynamic}, while its thermal amplitude remains around $60~\rm pm$  in air. The emergence of nonlinear behavior in higher modes of elastic structures and  Brownian motion in a nonlinear potential are both interesting questions requiring further study.

\begin{acknowledgments}
We acknowledge support from US NSF through grants CMMI-2001559, CMMI-1934271, CMMI-1934370, and CMMI-2001403.\textcolor{white}{ ............. ............. .............. .......... ........... ......... ..........}
\end{acknowledgments}


%


\end{document}